\documentclass[journal=jacsat,manuscript=article]{achemso}

\usepackage{chemformula} % Formula subscripts using \ch{}
\usepackage[T1]{fontenc} % Use modern font encodings
\usepackage{upgreek}
\author{Ali Panahpour}
\author{Jussi Kelavuori}
\author{Mikko Huttunen}
\email{ali.panahpour@tuni.fi}
\phone{+358504921272}
\affiliation[Unknown University]
{Photonics Laboratory, Physics Unit, Tampere University, FI-33014 Tampere, Finland}

\title[An \textsf{achemso} demo]
  {Purcell Effect in Epsilon-Near-Zero Microcavities}

\abbreviations{IR,NMR,UV}
\keywords{American Chemical Society, \LaTeX}

\begin{document}


\begin{abstract}
Epsilon-near-zero (ENZ) photonics presents a powerful platform for integrated photonic systems, enabling a range of novel and extraordinary functionalities. However, the practical implementation of ENZ-based systems is often constrained by high material losses and severe impedance mismatch, limiting the efficient interaction of light with ENZ media. To overcome these challenges, we introduce all-dielectric Bragg reflection microcavities operating at their cutoff frequency as a high-figure-of-merit ENZ resonant platform, providing an ultra-low-loss alternative for studying emission processes in ENZ media.
While Bragg cavities are well-established, their potential as ENZ resonant microcavities remains largely unexplored. We investigate the Purcell effect and quality factor in these structures, comparing their performance with those of the perfect-electric-conductor and metallic counterparts. Through analytical derivations based on Fermi’s golden rule and field quantization in lossless dispersive media, we establish scaling laws that distinguish these ENZ cavities from conventional resonators. Frequency-domain simulations validate our findings, demonstrating that in all-dielectric ENZ Bragg-reflection microcavities, the Purcell and quality factors scale as $L/\lambda_0$ and $(L/\lambda_0)^3$, respectively, where $L$ is the cavity length and $\lambda_0$ is the resonance wavelength.
Our results offer key insights into the design of ENZ-based photonic systems, paving the way for enhanced light-matter interactions in nonlinear optics and quantum photonics.

\end{abstract}

%\section{Introduction}
Epsilon-near-zero (ENZ) photonics provides a novel platform for developing new types of integrated photonic devices with unconventional functionalities \cite{Pollard2009,Engheta2013, Niu2018, Wu2021, Reshef2019, Kinsey2019, Lobet2023, Liberal2017,Hwang2023}. This is due to variety of extraordinary effects, offered by ENZ materials and structures, such as energy squeezing \cite{Alù2008}, tunneling through subwavelength channels and bends \cite{Silveirinha2006}, enhancement of optical nonlinearities \cite{Reshef2019, Fomra2024}, coupling of emitters for quantum entanglement \cite{Li2021,Issah2021} or quantum networks \cite{Vertchenko2019} and boosting of the Purcell factor (PF) in radiative systems \cite{Liberal2017, Alù2009, Alù2011, Fleury2013}. 
Conventional ENZ materials or nanostructures have been mainly based on conducting oxide or nitride materials as well as metamaterials consisting of metal--dielectric or semiconductor--dielectric components \cite{Niu2018, Wu2021}.
%Perfect electric conductor (PEC) waveguides, and more realistic but lossy 
Metallic waveguides, near their cutoff frequencies have also been utilized to emulate the properties of an ENZ medium \cite{Alù2011,Li2024}.

Despite known challenges, such as low transmission due to impedance mismatch with surroundings, ENZ media exhibit unique and highly advantageous properties. Notably, it has been shown that among the various subcategories of index-near zero (INZ) media, the PF diverges in the one-dimensional ENZ media, due to substantial growth of electromagnetic density of states (DOS) in such structures \cite{Lobet2020}. 
The ENZ-induced PF enhancement has been demonstrated both theoretically and experimentally in linear as well as nonlinear systems \cite{Vesseur2013, So2020, Caligiuri2018}. However, full exploitation of the Purcell enhancement of radiative processes is severely restricted by intrinsic losses \cite {Javani2016,Chebykin2015,Chebykin2016} which, as we will show, can prevent further enhancement of spontaneous emission rate  (SER) in ENZ media through coupling of embedded emitters to high-quality-factor ($Q$-factor) optical resonators. 
%In fact, as we will show, such couplings in metal or semiconductor structures can result in suppression of high-$Q$ resonances with no significant enhancement of radiative processes.

In this article, PF is investigated in high figure-of-merit (FoM) ENZ waveguides and cavities, which stand apart from conventional metal- or semiconductor-based structures, due to their exceptionally high ratio of the real to imaginary part of the effective permittivity. 
We explore PF in ideally lossless ENZ perfect electric conductor (PEC) as well as all-dielectric Bragg reflection waveguides (BRWs) and cavities near their cut-off frequencies, making comparisons with their metallic counterparts. 
Finally, scaling laws are derived for the PF and $Q$-factors of such ENZ structures, revealing a counter-intuitive trend of enhancement of PF by increasing the size of cavity.

The perturbative effect of  structured vacuum on radiative processes, known as Purcell effect was originally demonstrated by placing an emitter to a cavity \cite{purcell1946}. 
By tuning the emission wavelength of the emitter to the resonant mode of a nondispersive cavity, the emission is enhanced by a factor of 
\begin{equation}
 {F}_\mathrm{P}=\frac{6}{\pi^2}\left(\frac{\lambda}{2n_c}\right)^3\frac{Q}{V} \,,
\end{equation}
where $\lambda$ is the photon wavelength in vacuum, $n_c$ is the refractive index of medium inside the cavity, $Q$ is the $Q$-factor of the cavity and $V$ is the mode volume.
The conventional approaches for enhancing the PF are mainly based on three different classes of photonic structures: ultra-high-$Q$-factor (\(Q \,{\sim}\, 10^8\)) dielectric microcavities with %$Q$-factors in the range of \(10^8-10^9\) and 
$V$ much larger than \(\left(\lambda/2n_c\right)^3\) \cite{Kippenberg2004}, moderate $Q$-factor (\(Q \,{\sim}\, 10^4\)) photonic crystal nanocavities with mode volumes on the order of \(\left(\lambda/2n_c\right)^3\) \cite{Akahane2003}, and finally plasmonic nanoresonators with very small $V$ but rather small $Q$-factors (\(Q\le100\)).
Strongly dispersive cavities can also lead to enhanced $Q$-factor and Purcell effect \cite{Soljačić2005,Gao2016}.

Since a high FoM ENZ medium is inherently highly dispersive, it is expected that a resonant ENZ structure exhibits largely improved $Q$-factor and Purcell effect. The resonance of the cavity near the cutoff frequency of the structure can mitigate the challenge of weak coupling of light with ENZ media due to impedance mismatch. Furthermore, the elongation of wavelength within the ENZ region serves to alleviate scattering-based losses arising from structural imperfections \cite{Kippenberg2003}, and potentially can improve mode overlap and phase matching in nonlinear processes \cite{Suchowski2013,Fomra2024}.

%\section{Results and discussion}

To study the Purcell effect in ENZ structures in the weak interaction regime, we apply Fermi’s golden rule giving the SER of a quantum emitter: 
\begin{equation}
 \mathrm{\Gamma}=\frac{2\pi}{\hbar^2}\left|M_{12}\right|^2g(\omega) \,,
\end{equation}
in terms of transition matrix element \(M_{12}\) and optical DOS \(g(\omega)\). Applying the quantization procedure for a homogeneous and lossless dispersive medium~\cite{Milonni1995,Lobet2020}, the transition matrix element is given by:
\begin{equation}
 \left|M_{12}\right|^2=\frac{\hbar\omega}{2\varepsilon_0V}\frac{\left|\mathbf{p}\right|^2}{{n(\omega)n}_\text{g}(\omega)} \,,
\end{equation}
with $n$, $n_\text{g}$, %$V=L_x L_y L_z$ 
$V$ and $\mathbf{p}$ being the (real part of) medium phase index, group index, quantization volume and emitter's dipole moment, respectively. A straightforward method to realize low-dimensional (in $k$-space) ENZ structures is to use metallic waveguides with planar or rectangular geometries, operating near their cutoff frequencies. In the case of a one-dimensional rectangular waveguide with transverse dimensions of $L_x$ and $L_y$, and with idealized PEC walls, the DOS for an arbitrary quantization length of $L_z$ along $z$-axis is given by \(g\left(\omega\right)=\left(L_z/\pi c\right)n_\text{g}(\omega)\). Thus, by using Eqs.~(2) and (3), the SER of an emitter in the waveguide is given by: 
\begin{equation}
\Gamma^\mathrm{W}=\frac{\left| \mathbf{p} \right|^2}{ \varepsilon_0 \hbar }\frac{1}{ L_x L_y}\frac{\omega}{c}\frac{1}{n}  \,,
\end{equation}
and by normalizing it to the SER in free space, the PF turns into
\begin{equation}
F_\mathrm{P}^\mathrm{W}=\frac{3\pi}{L_x L_y}\frac{c^2}{ \omega^2}\frac{1}{n}  \,.
\end{equation}
The effective index of refraction in a PEC waveguide varies with wavelength as $n_{eff}(\lambda)=(1-\lambda/\lambda_c)^{1/2}$. Near the cutoff wavelength ($\lambda_c$), $n_{eff}$
becomes much smaller than the surrounding refractive index. For a finite-length waveguide, this index/impedance mismatch causes strong reflections at the waveguide ports. Therefore, the waveguide acts like a cavity, supporting only modes that satisfy the resonance condition $n_{eff}(\lambda)=m\lambda/2L_z$, where $m$ is an integer and $L_z$ is the waveguide length.
We focus on PF in this structure, particularly for the mode closest to cutoff. We assume that the cavity mode is tuned to the emission frequency (\(\omega_0\)) of a dipole emitter positioned at the center of the cavity.
%To study the PF in such stuctures, a single mode rectangular cavity is considered, consisting of a vacuum-filled PEC waveguide with transverse dimensions of $L_x$ and $L_y$, and a limited length of $L_z$. The resonance feedback is provided by reflections at the waveguide ports along $L_z$. We assume the cavity mode is tuned to the emission frequency (\(\omega_0\)) of a dipole emitter at the center of the cavity. 
The DOS around the resonant mode can then be written as a Lorentzian function \(g \left( \omega\right)=(2\delta \omega/\pi)/({4(\omega-\omega_0)}^2+{\delta \omega}^2)\), reducing at resonance to \(g\left(\omega_0\right) =(2/\pi)(Q/\omega_0)\). Here, \(Q=\omega_0/\delta \omega\) is the $Q$-factor of the cavity with the resonance width of \(\delta \omega\), which in terms of cavity finesse \(\mathcal{F}\), is given by  \(Q=\mathcal{F}L_z\omega_0/\pi c\). 
Since the cavity mode is near the cutoff frequency, it exhibits strong dispersion, modifying the free spectral range to \(c/2n_\text{g} L_z\) and the dispersive cavity $Q$-factor to \(Q_\text{disp}=Qn_\text{g}\) \cite{Soljačić2005,Gao2016}. The DOS is accordingly determined by \(g\left(\omega_0\right) =(2/\pi)(Q/\omega_0)n_\text{g} (\omega_0)\). Then applying Eqs.~(2) and (3), the spontaneous emission rate (SER) in the dispersive single-mode rectangular cavity is given by
\begin{equation}
\Gamma^\mathrm{C}=\frac{2\left| \mathbf{p} \right|^2}{ \hbar \varepsilon_0 V}\frac{Q}{n}  \,.
\end{equation}
After normalizing to the SER in vacuum, the PF in terms of \(\lambda_0=2\pi c/\omega_0\) %exactly at the resonance 
fulfills
\begin{equation}
{F}^\mathrm{C}_\mathrm{P}=\frac{3}{4\pi^2}\frac{\lambda_0^3}{V} \frac{Q}{n}\,.
\end{equation}
A more rigorous calculation of SER and PF in waveguides and cavities should account for the local variation of fields, by applying local density of states (LDOSs) and effective mode volume of the structures \cite{Benisty1999}. By implementing LDOSs, the SER will be proportional to the spatially varying electric field of the mode \(\mathbf{E}_m(\mathbf{r})\), normalized to its peak value, \(\Gamma( \mathbf{r})\propto \left|\mathbf{E}_m(\mathbf{r})\right|^2/\left|\mathbf{E}_m(\mathbf{r}_\mathrm{max})\right|^2\). Also, SER is inversely proportional to the effective mode volume defined as:
\begin{equation}
V_m=\frac{\int_{V} \varepsilon(\mathbf{r})\left|\mathbf{E}_m(\mathbf{r})\right|^2 \, \mathrm{d}\mathbf{r}}{\varepsilon(\mathbf{r}_\mathrm{max})\left|\mathbf{E}_m(\mathbf{r}_\mathrm{max})\right|^2}\, ,
\end{equation}
where, the integration is over the quantization volume, $V$. Assuming that the effective permittivity, $\varepsilon(\mathbf{r})$ is nearly invariant within the structure, the effective mode volume reduces to \(V_m=\int_{V} \left|\mathbf{E}_m(\mathbf{r})\right|^2 \, \mathrm{d}\mathbf{r}/\left|\mathbf{E}_m(\mathbf{r}_{max})\right|^2\). However, since we are interested in the average value of the SER and PF over the quantization volume, the average value of the spatially varying field will be balanced by the corresponding factor from the effective mode volume in the denominator. This ensures that the SER and PF relations (4)--(7) remain valid.
By comparing Eq.~(7) with Eq.~(1) for a cavity with fixed mode volume and reflectivity-limited $Q$-factor, we find that by inclusion of dispersion, the derived PF exceeds the conventional estimate by a factor of $f=n_c^3/n$.  For instance, with $n_c=1.5$ and $n=0.01$, the correction boosts the PF by 337.5 compared to the value given by Eq.(1).

Assuming \(L_y=\lambda_0/n_c \) (\(=2L_x/\sqrt{1+(L_x/L_z)^2}\)), \(L_x=\lambda_c/2n_c\) ($\approx\lambda_0/2n_c$ for $L_z\gg L_x$), \(L_z=\lambda_0/2n\) and using \(Q=(2n_cL_z/\lambda_0)\mathcal{F}\), we see from Eq.~(7) that PF is proportional to \(\mathcal{F}/n\). In the absence of material losses, the cavity finesse $\mathcal{F}$ is determined by the reflectivity finesse, \(\mathcal{F}_R=\pi\sqrt{R}/(1-R)\) in terms of reflectivity of the ports $R$. For ENZ/INZ cavities, the reflectivity finesse is approximately \(\mathcal{F}_R\approx\pi/4n\) (see Supporting Information). 
Therefore, the PF and $Q$-factor of the dispersive cavity are given by

\begin{equation}
{F}^\mathrm{C}_\mathrm{P}\approx\frac{3}{4\pi} \frac{n_c^3}{n^2},
\end{equation}
\begin{equation}
{Q_\text{disp.}=Qn_g\approx\frac{\pi}{4} \frac{n_c}{n^3}}.\, 
\end{equation}
Then by using the fundamental cavity resonance condition \(n= \lambda_0/2L_z\), we find \({F}^\mathrm{C}_\mathrm{P} \propto(L_z/\lambda_0)^2\) and \(Q_\text{disp.} \propto(L_z/\lambda_0)^3\). These scaling laws are in contrast to the cases of ordinary microcavities, where PF is reduced by increasing the mode volume.

In practice, the divergence of PF at zero effective index of refraction, as predicted by Eqs.~(5) and (7), is prevented by material losses. The detrimental impact of material losses on PF is much more pronounced in resonant cavities due to multiple passes of light inside the cavity. To demonstrate this  effect in metallic cavities, we conducted COMSOL simulations exploiting the mathematical analogy of quantum and classical treatments of the SER. According to Poynting’s theorem the radiated power of any current distribution \(\mathbf{j}\)(\(\mathbf{r})\) with a harmonic time dependence in a linear medium must be equal to the rate of energy dissipation given by \(\mathrm{d}W/\mathrm{d}t=-1/2\int_{V} \mathrm{Re}\left\{\mathbf{j}^\ast \cdot\mathbf{E}\right\} \mathrm{d}V\). For a current distribution corresponding to a point dipole emitter \(\mathbf{j}\)(\(\mathbf{r})=-\mathrm{i}\omega \mathbf{p} \delta (\mathbf{r}-\mathbf{r}_0)\) with dipole moment \(\mathbf{p}\) located at \(\mathbf{r}_0\), the equation can be rewritten in terms of the electric dyadic Green function $\mathbf{G}$, as \(\mathrm{d}W/\mathrm{d}t=\omega^3/2c^2\varepsilon_0 \varepsilon \left[ \mathrm{Im}\, \left\{\mathbf{p} \cdot\mathbf{G}(\mathbf{r}_0,\mathbf{r}_0;\omega) \cdot \mathbf{p}\right\} \right]\), which establishes a link between the quantum and classical formalisms for PF calculations \cite{Novotny2012}. Following this approach, we place a gold nanoparticle carrying a harmonic current density, acting therefore as an oscillating dipole emitter, to the center of the cavity. Then we compute the power dissipation and normalize it to its corresponding value in free space, allowing us to determine the PF within the structures \cite{Chen2010}.

Calculated PF curves for cavities formed by either PEC (solid curve) or gold (dashed curve) walls are shown in Fig.~1. The cavity dimensions are \(L_x=0.5 \;\upmu\)m, \(L_y=1\;\upmu\)m, and \(L_z=8 \;\upmu\)m. The dipole moment of the emitter is along $y$-axis, exciting the lowest order TE mode of the cavity. For the case of gold cavity, $L_x$ is reduced by the value of skin depth to keep the cutoff at \(\lambda_\text{c}=1\;\upmu\)m. %(the effective thickness of the waveguide is larger than the distance between the walls due to some penetration of waves into metallic walls). 
The resonance feedback is provided by impedance mismatch through inclusion of two waveguide sections filled with materials of refractive index \(n=2\) at the cavity ports.

\begin{figure}
  \includegraphics[scale=0.53]
  {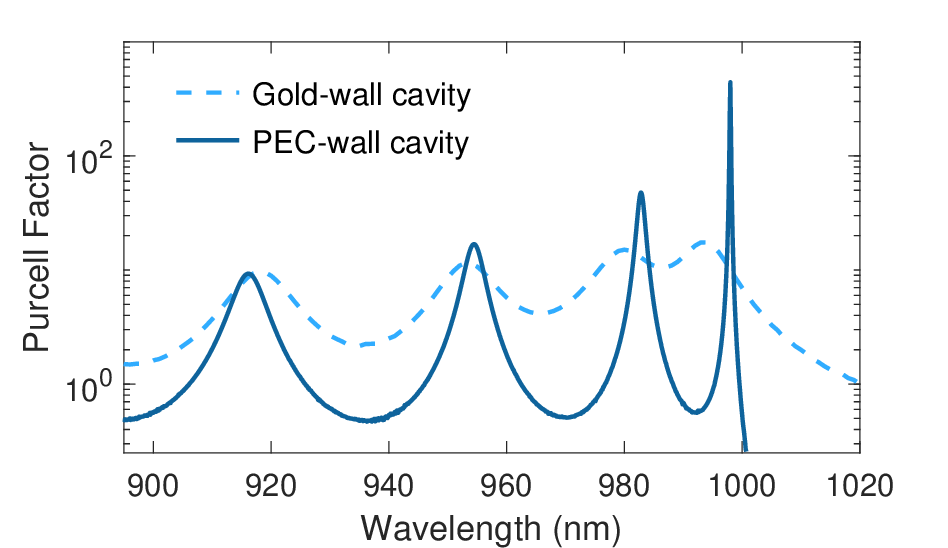}
  \caption{Numerically calculated PFs for cavities with PEC (solid curve) and gold (dashed curve) walls and cutoff wavelength of \(\lambda_\text{c}=1\;\upmu\)m.}
  \label{fgr:example}
\end{figure}

The curves show that the free spectral range and spectral width of the resonances are reduced in both PEC and gold-wall resonators, when approaching the cutoff wavelength. This is due to increased dispersion inside the cavities near the cutoff wavelength. However, the $Q$-factor and PF of the gold cavity are considerably lower than those of the PEC cavity, due to the ohmic losses of gold.
The losses also adversely affect the dispersion profile of modes in a waveguide. As depicted in  Fig.~2a, the dispersion of the real part of effective index \(n(\omega)\), and consequently $n_\text{g}$ %or \(n^{-1}(\omega)\)
in a PEC waveguide tends to infinity when approaching the cutoff wavelength of \(\lambda_c=1\;\upmu\)m. 
However, for a gold-wall waveguide (Fig.~2b), 
%there is a notable increase in dispersion and $n_\text{g}$ around the cutoff wavelength, 
the group index is maximized already before the cutoff and gradually tends to zero at longer wavelengths accompanied by a substantial rise in imaginary part of effective index $\kappa$, responsible for propagation losses.

\begin{figure}
  \includegraphics[scale=0.56]
  {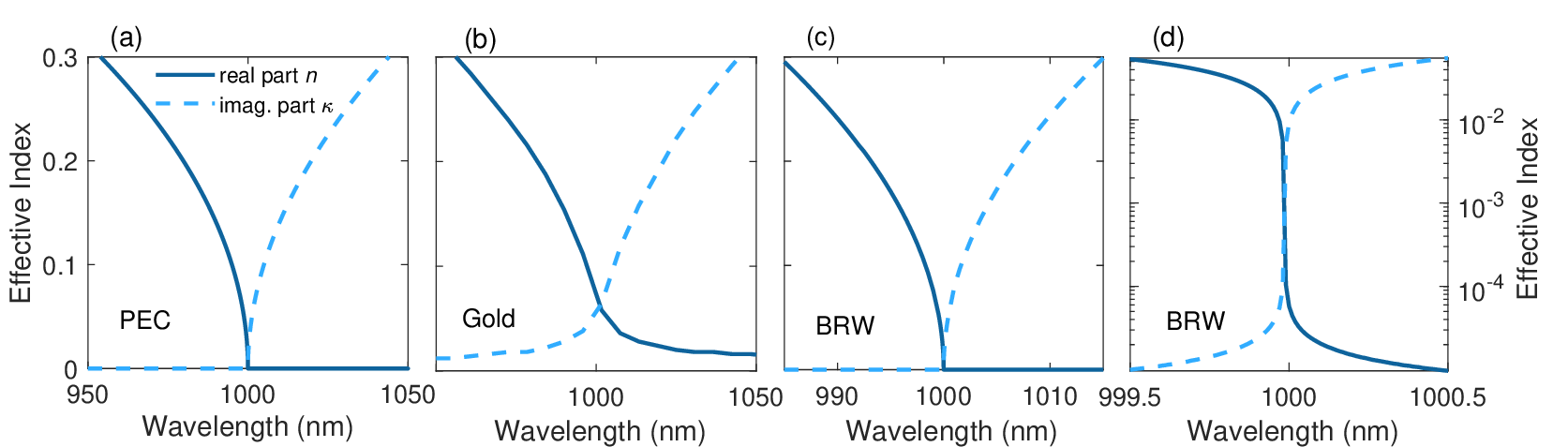}
\caption{\label{fig:2} Dispersion curves of (a) a PEC waveguide, (b) a gold-wall waveguide and (c) a BRW around the cutoff wavelength of \(\lambda_\text{c}=1\;\upmu\)m. The transverse dimensions of the PEC and gold wall waveguides are the same as in Fig.~1.
 (d) A magnified view of the BRW dispersion curve near the cutoff wavelength, shown in logarithmic scale.} 
\end{figure}

Three-dimensional (3D) COMSOL simulations were also conducted to study the dependence of PF and $Q$-factor (blue dots in Figs.~3a and 3b) on the effective index of refraction (\(n= \lambda_0/2L_z\)) in a PEC-wall cavity, through changing the cavity length \(L_z\). The cavity core and its surroundings are vacuum, with resonance feedback arising from reflections at the cavity ports due to impedance mismatch. The straight lines in Fig.~3, correspond to different functional dependence %of PF and $Q$-factor 
on $1/n$. The results confirm the scaling of PF and $Q$-factor as $n^{-2}$ and $n^{-3}$, respectively.

\begin{figure}
  \includegraphics[scale=0.6]
  {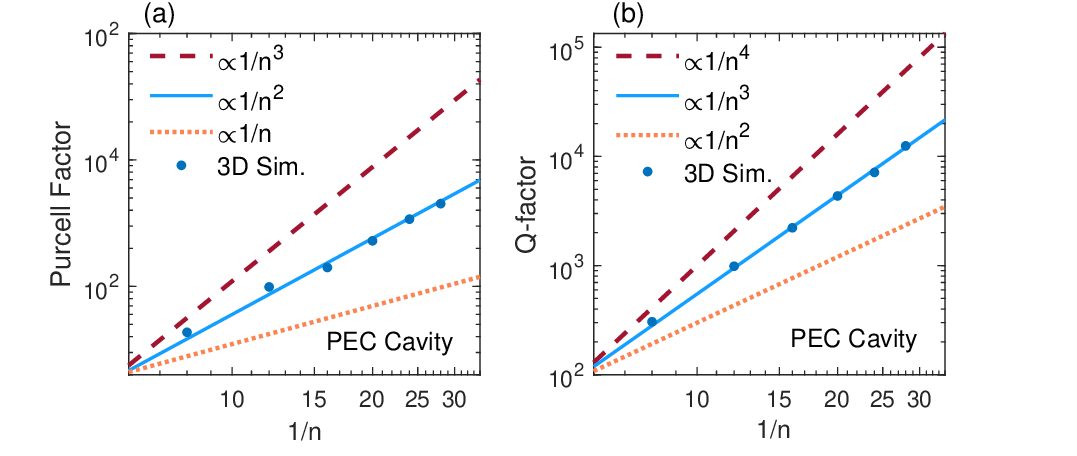}
\caption{\label{fig:3} Simulation results (blue dots) showing (a) PF and (b) $Q$-factor values in PEC-wall cavities with $L_x=0.5\;\upmu $m, $L_y=1~\upmu$m, and $L_z=4, 6, 8, 10, 12, 14\;\upmu$m, corresponding to $1/n\approx 8, 12, 16, 20, 24$ and $28$, respectively. The lines represent different dependence of PF and $Q$-factor on $1/n$.
 }
\end{figure}

The computed PF values correspond to the center of the cavity. To compare them with the average PF values predicted by Eq.~(9), the simulation data should be scaled by a factor of 1/4, as the effective cavity mode volume for the lowest order TE mode is a quarter of the geometrical volume used in Eqs.~(7) and (9)\cite{Benisty1999}. 
The simulation results for the PF and $Q$-factor are nearly 1.5 times higher than those predicted by Eq.~(9). This discrepancy can be attributed to the inaccuracy of the reflectivity expression used at the cavity ports, given by \(R=(n-1/n+1)^2\). While this relation holds for an infinitely extended boundary, the finite size of the cavity ports—comparable to the wavelength—introduces diffraction effects that deviate from this idealized model \cite{Yee1968,Li2016}. Even a small variation in reflectivity can lead to significant changes in the $Q$-factor and consequently the PF.

In practice, perfectly lossless counterparts of PEC waveguides and cavities at optical frequencies do not exist. However, all-dielectric planar BRWs \cite{Yeh1976} operating near cutoff can exhibit low-loss ENZ behavior, featuring strong dispersion and a high group index ($n_\text{g}$) quite similar to PEC waveguides (Figs.~2a and 2c).
This enhancement of $n_\text{g}$ has been studied in near-cutoff BRWs, focusing on the associated slow-light features \cite{Kozlov2010,Fuchida2012}. 
%Orders of magnitude enhancement of nonlinear effects such as Raman scattering has been reported in semiconductor half-wave Bragg cavities.
Earlier work has also shown the possibility of controlling the intensity, rate or angular distribution of spontaneous emission in Bragg reflector (BR) microcavities \cite{Björk1991,Deppe1991,Vredenberg1993}. The above approaches have been based on Fermi’s golden rule or first-order perturbation theory of coupled electronic-electromagnetic systems in multimode cavities with virtually infinite lateral dimensions. Here, we study the PF and $Q$-factor in single-mode near-cutoff BR cavities, with laterally confined dimensions (Fig. 4a), where the cavity mode frequency is set by the structure’s finite size. The theory is based on Fermi's golden rule and quantization of lossless dispersive media. The results are represented in terms of effective refractive index of the cavity, which holds particular relevance in the context of ENZ/INZ structures. 
%We also note that the studied all-dielectric ENZ microcavities and the associated waveguiding mechanism are quite different from the so-called ENZ-insulator-ENZ waveguide structures, where the mode confinement is due to total reflection of waves at the insulator-ENZ interfaces.

\begin{figure}
  \includegraphics[scale=0.35]
  {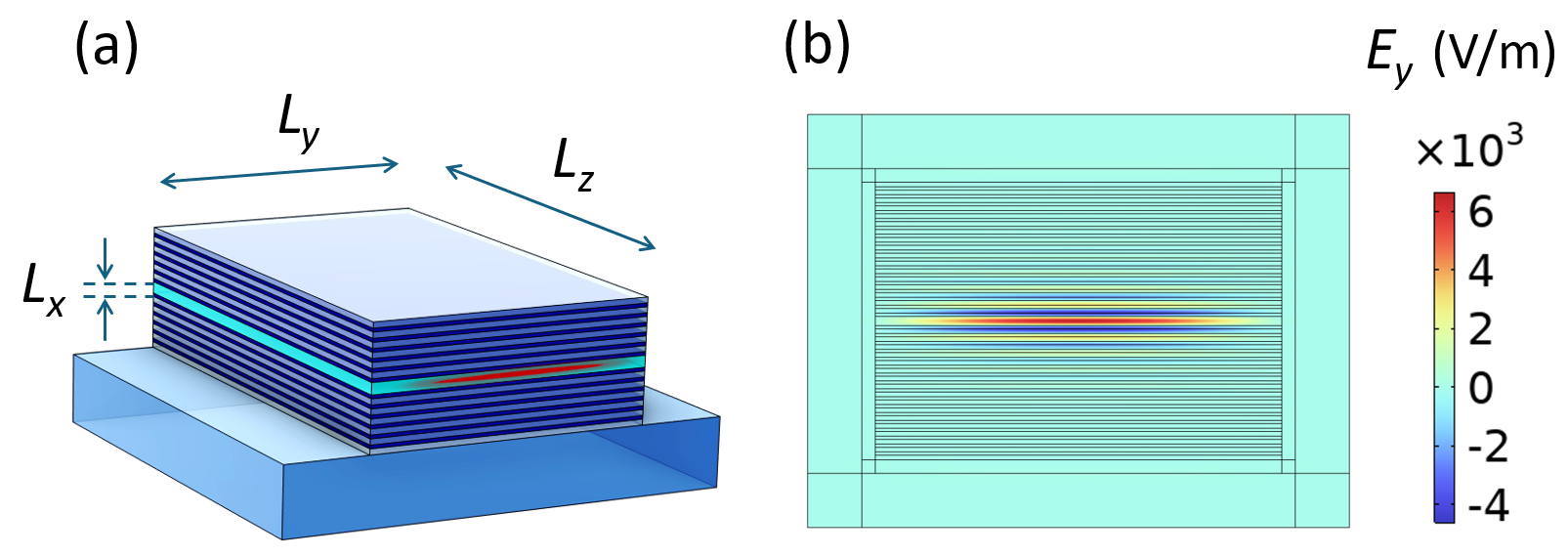}
\caption{\label{fig:3} (a) Schematic view of a BR  microcavity and (b) its transverse cross-section view, showing the $y$-component of electric field, corresponding to an ENZ mode of the waveguide, computed by 2D mode-analysis simulation. }
 
\end{figure}

The guided waves in BRW structures propagate parallel to the Bragg walls with very little losses compared to common ENZ structures. In principle, the radiative losses through the Bragg walls can be arbitrarily reduced by increasing the number of cladding layers. Therefore, the supported low-loss ENZ modes show strong dispersion and large group indices (\(n_\text{g}\approx n^{-1}\)) near the cutoff frequencies \cite{Nistad2006,Xu2002}, behaving thus very similar to PEC waveguides (Figs.~2a and 2c). Fig. 2d is a magnified view of the dispersion curve in Fig. 2c around the cutoff wavelength, shown in logarithmic scale.
The dispersion curves in  Fig.~2c are obtained by two-dimensional (2D) COMSOL mode-analysis simulation of a planar quarter-wave BRW with infinite lateral dimensions. The electric field polarization of the mode is parallel to the Bragg reflectors. The waveguide core material is assumed to be borosilicate glass while the cladding structures on each side consisted of 20 alternating pairs of borosilicate and TiO$_2$ layers, with phase thicknesses of \(\pi/2\) and refractive indices of $n_\mathrm{BK7}=1.51+\mathrm{i} 9.93\times10^{-9}$ \cite{schott} and $n_{\mathrm{TiO}_2}=2.31+\mathrm{i} 10^{-6}$ \cite{Zhukovsky2015}. The magnified view of the BRW dispersion curve in Fig.~2d shows that the FoM of the waveguide at \(\lambda=999.5\)~nm is as high as $\epsilon'/\epsilon''=n^2-\kappa^2/2n\kappa\approx2677$, with $n=0.053$ and $\kappa=9.9\times10^{-6}$ as the real and imaginary parts of the effective index. 

In BRWs with virtually infinite lateral dimensions, a continuous range of propagating modes exists for wavelengths below the cutoff. These modes can be excited by illuminating the structure either at normal incidence or at an oblique angle (with respect to the $x$-axis in Fig.~4a). Normal excitation primarily excites the mode closest to cutoff frequency, while increasing the incident angle allows excitation of other waveguide modes.
These propagating modes correspond to resonant modes of the structure along the $x$-axis, effectively forming a half-wave BR cavity. The resonances give rise to sharp transmission peaks within the stop-band of the structure, as shown in Fig.~5, corresponding to different number of cladding layers. The figure illustrates the transmittance of a structure, with the same parameters as Fig.~2c, under normal illumination along the $x$-axis. As the incident angle increases, the resonant transmission peaks blue-shift from the cutoff wavelength, reflecting the excitation of other horizontally propagating waveguide modes.

\begin{figure}
  \includegraphics[scale=0.45]
  {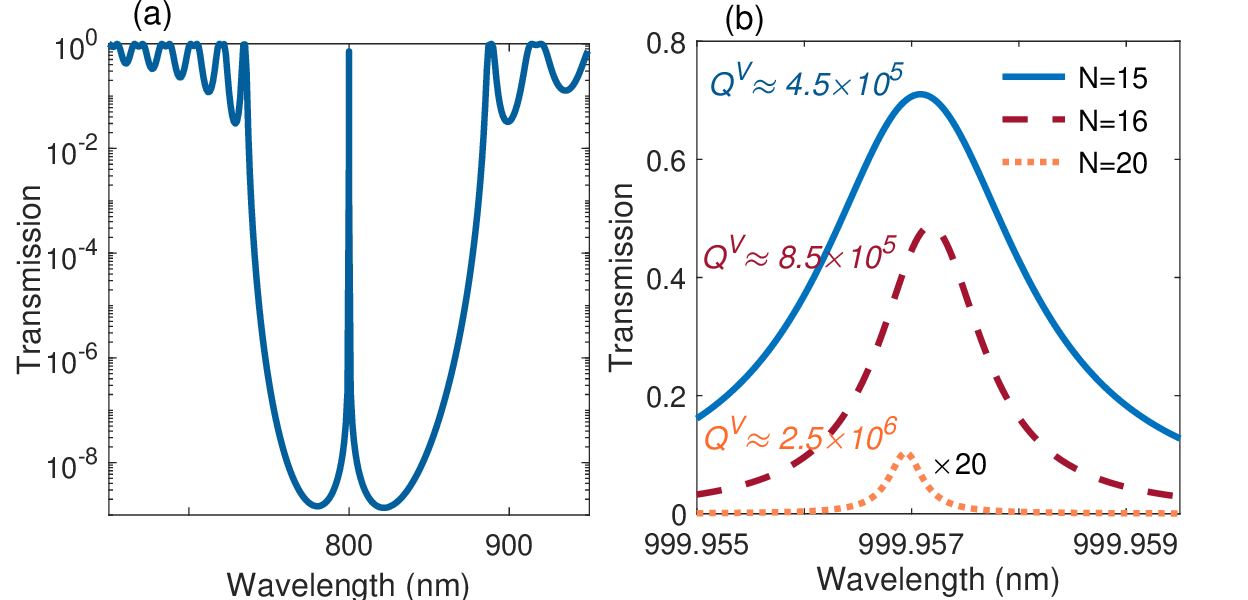}
\caption{\label{fig:5} Normal incidence transmittance through the BRW with parameters as in Fig.~2c, with electric field polarization parallel to the Bragg walls. (a) Transmittance corresponding to the number of cladding pairs of $N=15$. (b) a zoomed view of transmission peaks for $N = 15$, 16 and 20, represented by solid, dashed and dotted curves, respectively. }
\end{figure}

In contrast, finite-length BRWs near cutoff exhibit strong reflections from the waveguide ports, leading to cavity-like behavior along the resonator axis ($z$-axis in Fig.~4). As a result, only a discrete set of BRW modes can propagate and resonate along the cavity axis. The resonant modes are defined by the length of the cavity along $z$. Coupling light into such near-cutoff cavities through the ports is challenging due to significant impedance mismatch with the surroundings. Instead, illumination close to normal incidence provides an effective way to excite the cavity modes, with each mode corresponding to a specific incident angle of the illuminating wave.

The length $L_z$, determines both the mode frequency and the effective index. As $L_z$ increases, the effective index approaches zero, leading to an enhancement in PF, as predicted by Eq.~(7). However, the PF cannot be enhanced indefinitely in this manner, as increasing the cavity length and approaching the cutoff wavelength leads to a rapid decline in the FoM, as evident in Fig.~2d. Moreover, maintaining high FoMs while achieving near-zero effective indices is constrained by the cavity width ($L_y$). Our mode analysis simulations indicate that reducing the cavity width increases lateral radiative losses, thereby lowering the FoM. Conversely, increasing $L_y$ mitigates radiative losses along the $y$-axis. Specifically, for a BRW with the same parameters as in Fig. 2c, but with limited width of $L_y=10$~$\upmu$m, simulations yield an effective index of $n=0.047+\mathrm{i}1.29\times10^{-5}$, at \(\lambda=999.5\)~nm, corresponding to FoM of $1828$. Expanding the width to $15$~$\upmu$m ($20$~$\upmu$m), raises the FoM to 2483 (2621), approaching the upper limit of 2677 defined in Fig. 2c for an infinitely wide waveguide (modeled using periodic boundary conditions on the lateral boundaries). This means that the dispersion curves of laterally infinite BRWs set the upper bound for the FoM in BR cavities. Thus, as the cavity length increases and the effective index decreases, preserving the waveguide’s FoM necessitates a proportional increase in $L_y$ relative to $L_z$. This contrasts with PEC structures, where the effective index can be reduced arbitrarily by extending the cavity length, leading to a distinct scaling behavior of the PF with respect to $1/n$ in BRW cavities compared to PEC cavities.

To estimate the PF in a typical BR cavity shown in Fig. 4a, we use Eq.~(7) and set \(L_x =\lambda_c/2n_c\) (in terms of cutoff wavelength and real part of core index, $n_c$) which is slightly larger than \(\lambda_0/2n_c\). We then choose \(L_z=\lambda_0/2n\) and \(L_y=\lambda_0/4n\), yielding \({F}^\mathrm{BRC}_\mathrm{P}\approx (12/\pi^2)nn_cQ\). By substituting $Q\approx\pi n_c/4n^2$ for near-zero $n$, and incorporating local-field effects, we obtain:
\begin{equation}
{F}^\mathrm{BRC}_\mathrm{P}\approx\frac{3 L^2}{\pi}\frac{n_c^2}{n} \,,
\end{equation}
where $L$ is the Lorentz--Lorenz local-field correction factor, \(L=(2+\varepsilon)/3\) for an atom inside the core~\cite{Milonni1995,Dexter1956}. In the ENZ regime where the effective epsilon tends to zero ($\varepsilon\rightarrow0$), this factor reduces to $L\approx2/3$. So, for the BR cavity with core index of $n_\mathrm{core}=1.51$, we obtain ${F}^\mathrm{BRC}_\mathrm{P}\approx0.97\times n^{-1}$ that evaluates to ${F}^\mathrm{BRC}_\mathrm{P}\approx97$ for $n=0.01$.
The refractive index $n=0.01$ near the cutoff wavelength $\lambda_c=1000$~nm, corresponds to the cavity dimensions of $L_z=\lambda_0/2n\approx50~\upmu$m and $L_y=25~\upmu$m. Mode-analysis simulations of a BR cavity with these parameters reveal that $n=0.01$ corresponds to an imaginary part of $k=4.99\times10^{-5}$  at the resonance wavelength $\lambda_0=999.6$~nm, yielding a FoM of 200.4. Furthermore, the simulated resonance wavelength aligns well with the analytical expression \(\lambda_0=2n_cL_x/\sqrt{1+(L_x/L_z)^2}\).

To validate Eq.~(11) numerically, full-wave 3D simulation of BR cavities shown in Fig.~4a is required. However, COMSOL simulation of such structures with dimensions much larger than the wavelength resulted in significant convergence issues. These challenges arose from the large simulation domain and the high number of quarter-wave Bragg layers, which necessitated an extremely fine mesh. To overcome this and study the dependence of PF/$Q$-factor on the near-zero (NZ) cavity index, we replaced the BRs in COMSOL simulations with highly reflective thin-film virtual mirrors (VMs); see Supporting Information.
We assigned the VM with an NZ refractive index to avoid fine-mesh requirements in simulations. To mimic BRs, we also assigned the VM with an NZ-impedance to exhibit high reflectance. Properly tuning the permittivity and permeability of the VM allows its reflectance to match that of real BRs.
To further facilitate simulation convergence and reduce computation time, we considered cavities with lower $Q$-factors, on the order of $10^5$. In the Supp. Info. document, it is shown that a typical laterally infinite cavity with SiO$_2$ core and Bragg reflectors each composed of 16 pairs of quarter-wave layers exhibits a Q-factor of $\approx2.78\times10^5$ along $x$-axis. By employing a $40$~nm thin-film VM with permittivity $\varepsilon_{\mathrm{VM}}=3.65\times10^3$ and permeability $\mu_{\mathrm{VM}}=10^{-5}$, we achieved reflectivities closely matching those of the BRs in the spectral range of interest. As demonstrated in Supp. Info., a Fabry--Pérot cavity with SiO$_2$ core sandwiched between these VMs closely replicates the mode and transmission profiles as well as $Q$-factor of the BR cavity.
We computed the PF/$Q$-factor of the same VM cavity but with several finite lengths and the same size ratio of $L_z/L_y=2$ used to derive Eq.~(11). The results, expressed in terms of the inverse of the NZ effective index, are shown in Fig.~6. 
In Fig.~6a, the blue dots denote the simulation results, the solid blue line corresponds to the function PF$\approx7.14\times(1/n)$ predicted by Eq.~(11) and the dashed line represents a function with an inverse square dependence on $1/n$.
The simulation data correspond to the PF at the cavity center, whereas Eq.~(11) provides the PF averaged over the cavity volume. The numerically calculated ratio of the mode volume to cavity volume is nearly equal to 1/8. Therefore, the values given by Eq.~(11) were scaled by a factor of 8 to represent the PF at the center of the cavity, shown as PF$\approx7.14\times(1/n)$ in Fig.~6a.

The cubic dependence of the Q-factor on $1/n$, is clearly confirmed in  Fig. 6b, similar to the case of PEC ENZ cavity. 
Although the cavity is not lossless (the core refractive index is complex; $n_c=1.4505+i10^{-6}$), the simulation data in Figs.~6a an 6b, nearly follow the analytical predictions derived for lossless cavities, scaling as $7.14\times(1/n)$ for the PF and $1.14\times(1/n^3)$ for the $Q$-factor.

\begin{figure}
  \includegraphics[scale=0.6]
  {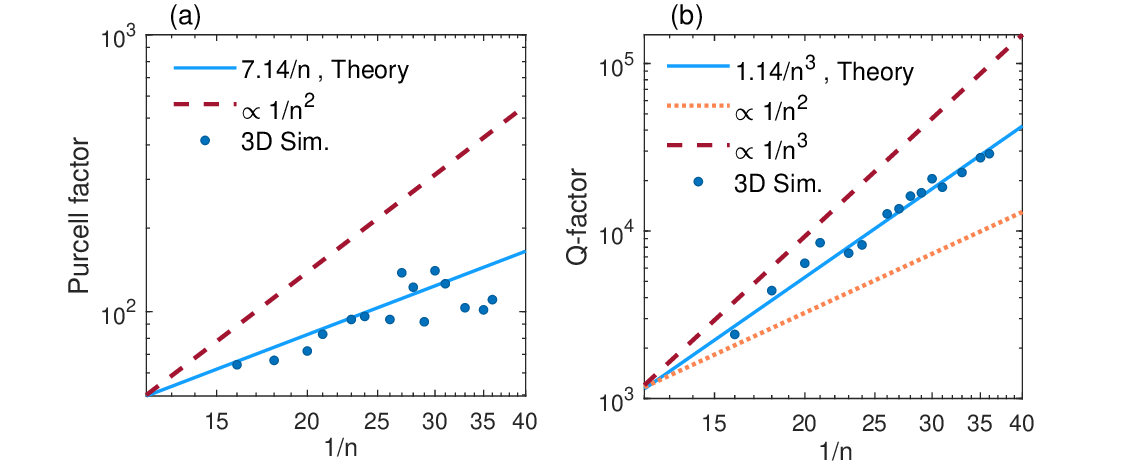}
\caption{\label{fig:5} Simulation results (blue dots) showing (a) PF and (b) $Q$-factor values in VM cavities with cutoff wavelength of $\lambda_c=1000$~nm, $L_x=\lambda_c/2n_{\mathrm{core}}$ (half-wave thickness), $L_y=L_z/2$, and several values of $L_z$ from $8$ to $18\;\upmu$m, corresponding to $1/n\approx 16-36$. The lines represent different proportionality of PF and $Q$-factor on $1/n$. }
\end{figure}

The cavity losses primarily arise from material absorption, along with two additional loss mechanisms: leakage through the BRs/VMs in the vertical (V) direction and through the cavity ports or sidewalls in the horizontal (H) direction. Consequently, the overall output coupling $Q$-factor is given by \(1/Q=1/Q^\mathrm{V}+1/Q^\mathrm{H}\) with $Q^\mathrm{H}\approx\pi n_c/4n^2$,
%For smaller cavities, where $Q^\mathrm{H}\approx\pi n_c/4n^2$ is significantly lower than $Q^\mathrm{V}$, increasing the cavity length enhances reflections from the cavity ports, causing the overall $Q$-factor to follow the $1/n^3$ scaling. However, as $Q^\mathrm{H}$ approaches $Q^\mathrm{V}$, both the quality and the Purcell factors deviate from their respective $1/n^3$ and $1/n$ trends. 
and by using \({F}^\mathrm{BRC}_\mathrm{P}\approx (12/\pi^2)nn_cQ\), the PF can be written as
\begin{equation}
{F}^\mathrm{BRC}_\mathrm{P}\approx \frac{12nn_c}{\pi^2}\frac{Q_\mathrm{H}}{1+Q_\mathrm{H}/Q_\mathrm{V}} \,,
\end{equation}
which for \(Q_\mathrm{V}\to \infty\) reduces to Eq.~(11) -- regardless of the local-field correction factor -- corresponding to the case of a PEC cavity. 
For smaller cavities, where $Q^\mathrm{H}$ is significantly lower than $Q^\mathrm{V}$, the PF and $Q$-factor follow the rules in (9) and (10). However, as $Q^\mathrm{H}$ approaches $Q^\mathrm{V}$, both the quality and Purcell factors deviate from the corresponding Eqs.~(9) and (10). 

Therefore, the upper limit of the $Q$-factor for BR cavity is ultimately set by $Q^\mathrm{V}$, which is primarily constrained by losses associated with the Bragg mirrors. The BR cavity losses can be minimized by applying low-loss, high-purity core and clad materials, super-polished substrates to reduce scattering, and a large number of cladding layers to suppress radiative leakage. By employing these techniques, BR cavities with exceptionally high finesse values on the order of  $10^5-10^6$ have been demonstrated \cite{Hunger2010, Rempe1992}.
Therefore, in practice, PF in BR cavities is restricted by feasibility of achieving high-reflectivity Bragg mirrors.
%$Q$-factors in BRW structures, mainly affected by material losses and structural imperfections. 
While commercial dielectric super-mirrors are available with reflectivities exceeding $99.998\%$,  
reflectivities of micro-scale structures and consequently the cavity $Q$-factor are known to decrease due to limited lateral size and scattering from sidewalls. However, very high reflectivities can be achieved already in structures with lateral dimensions from a few microns to a few tens of microns \cite{Reitzenstein2007,Haisler2004}.   
% Lei, C. et al, 1991, J. of App. Phys., 69(11), 7430-7434, Haisler, V. A. et al, 2004, J. of App. Phys., 96(3), 1289-1292

%Finally, we note that for experimental characterization of the PF, the emission linewidth of emitters should be narrower than the cavity resonance linewidth. This could be achieved by carefully selecting the emitter and minimizing linewidth broadening mechanisms, for example by operating at cryogenic temperatures or using crystalline structures instead of amorphous materials. 
In summary, we have derived scaling laws for the quality and Purcell factors in lossless or ultra-low-loss ENZ microcavities. For PEC-walled ENZ rectangular cavities, these factors scale respectively as $n^{-3}$ and $n^{-2}$ with the effective refractive index. In all-dielectric ENZ BR cavities near cutoff, they follow  \(n^{-3}\) and $n^{-1}$, respectively. However, for large cavities, the upper limit of these factors is ultimately constrained by Bragg reflector reflectivity. We think that ENZ microcavities offer exciting opportunities for low-threshold lasing, phase-match-free nonlinear processes, and studies of strong light-matter interactions.

\begin{acknowledgement}

All authors acknowledge the support of the Flagship of Photonics Research and Innovation (PREIN) funded by the Research Council of Finland (grant no.~320165). JK also acknowledges the Magnus Ehrnrooth foundation for their PhD grant.

\end{acknowledgement}

\begin{suppinfo}

The supporting information includes an estimation of the reflectivity finesse for an ENZ cavity based on the effective NZ refractive index. It also discusses the similarity between the transmittance and mode profiles of VM-based and BR cavities.

\end{suppinfo}

\bibliography{achemso-demo}

%\section*{Abstract}
\section*{Supplementary material}
This document provides supplementary information to "Purcell Effect in Epsilon-Near-Zero Microcavities"

\section{Estimation of reflectivity finesse of ENZ microcavities in terms of effective index of refraction}

In the Bragg reflection or PEC cavities described in the main text, the resonance feedback along $z$ direction is provided by partial reflections from the cavity ports.
The reflections depend on the contrast between the characteristic impedance of the waveguide, $Z_w=\sqrt{\mu/\epsilon}$ in terms of the effective permeability and permittivity of the waveguide and the impedance of the surrounding medium, $Z_s$. The reflection coefficient at the ports is given by: 
\begin{equation}
R = \left( \frac{Z_s - Z_w}{Z_s + Z_w} \right)^2 
\,.
\end{equation}
Using the relations $Z_s=Z_0/n_s$  and $Z_w=Z_0/n_\mathrm{eff}$, where the impedances are expressed in terms of the free-space impedance and the effective refractive indices of the waveguide and surrounding medium, the reflectivity simplifies to: 
\begin{equation}
R = \left( \frac{n_\mathrm{eff} - n_s}{n_\mathrm{eff} + n_s} \right)^2 
\,.
\end{equation}
In the ENZ regime when $n_\mathrm{eff}$ tends to zero, the reflectivity approaches the unity. Thus, the output coupling from the cavity ports is strongly influenced by the cavity effective index. Regarding the reflectivity finesse  \(\mathcal{F}_R=\pi\sqrt{R}/(1-R)\) of the cavity with symmetrical ports along the $z$-axis, in the limit of $n_\mathrm{eff}\rightarrow 0$, applying relation (2) and simplifying the expression, we arrive at:
\begin{equation}
\mathcal{F}_R=\frac{\pi}{4}\frac{n_s}{n_\mathrm{eff}}\propto \frac{1}{n_\mathrm{eff}}
\,.
\end{equation}
Our simulation results in the manuscript's Fig.~3 and 6 confirm the validity of this approximation used for calculation of Purcell and quality factors. 

\section{Implementing thin-film virtual mirrors instead of Bragg mirrors in COMSOL simulations of 3D Bragg reflector cavities}

In this section we compare the performance of a typical BR cavity with a Fabry Perot cavity consisting of carefully designed thin film virtual mirrors (VMs). We show that VMs can serve as an effective alternative to BRs in 3D COMSOL simulations of BR cavities, helping to resolve convergence issues. We consider a BR cavity consisting of a half-wave-thickness SiO$_2$ core sandwiched between two cladding with 16 pairs of quarter-wave SiO$_2$ and Ta$_2$O$_5$ layers. The real parts of the materials' refractive indices are $n_{SiO_2}=1.4504$ [S1] %\cite{Malitson1965} 
and $n_{Ta_2O_5}=2.0990$ [S2] at the cutoff wavelength $\lambda_c=1000~$nm of the structure. The imaginary part of the refractive indices is assumed to be  $\kappa\sim10^{-6}$. This estimation is based on the experimental data in [S3] and reproducing their results by tuning the $\kappa$ in COMSOL simulation of the same Bragg mirror. 

%The transmission peak of the structure is at $\lambda\approx999.971~nm$, with Q-factor of $Q\approx3.4\times10^5$.

%We note that 3D COMSOL simulation of such BR cavities with dimensions much larger than the wavelength poses critical convergence issues. This is due to both the large size of simulation domain and the large number of quarter-wave Bragg layers, which require very fine mesh structure. Therefore, to overcome this issue and study the functional dependence of PF/$Q$-factor on the near-zero (NZ) index of cavity, we replace the BRs in COMSOL simulations, with highly reflecting thin-film virtual mirrors (VMs).  
We assume the refractive index of the VMs to be near zero, minimizing fine-mesh requirements in simulations while ensuring high reflectance comparable to Bragg reflectors by assigning them a near-zero impedance. By appropriately selecting the relative permittivity and permeability of the VMs, we can precisely match their reflectance to that of actual BRs.

By employing a $40$~nm thin-film VM with permittivity $\varepsilon_{VM}=3.65\times10^3$ and permeability $\mu_{\mathrm{VM}}=10^{-5}$, we achieve reflectivities closely matching those of the SiO$_2$/Ta$_2$O$_5$ Bragg cavity in the spectral range of interest. Notably, the reflectivity of VMs does not need to be an exact match to that of BRs, as the simulated cavities are relatively small. Consequently, the reflectivity from the cavity ports is significantly lower than that of the BRs, making the Purcell factor or $Q$-factor primarily dependent on the ports' reflectivity rather than the BR reflectivities.

Considering a half-wave SiO$_2$ core between the two VMs and illuminating the cavity with a normally incident and horizontally polarized wave, we obtain the mode profile shown in Fig. 1a, which closely resembles that of the BR cavity in Fig. 1b. 

\begin{figure}
  \includegraphics[scale=0.41]
  {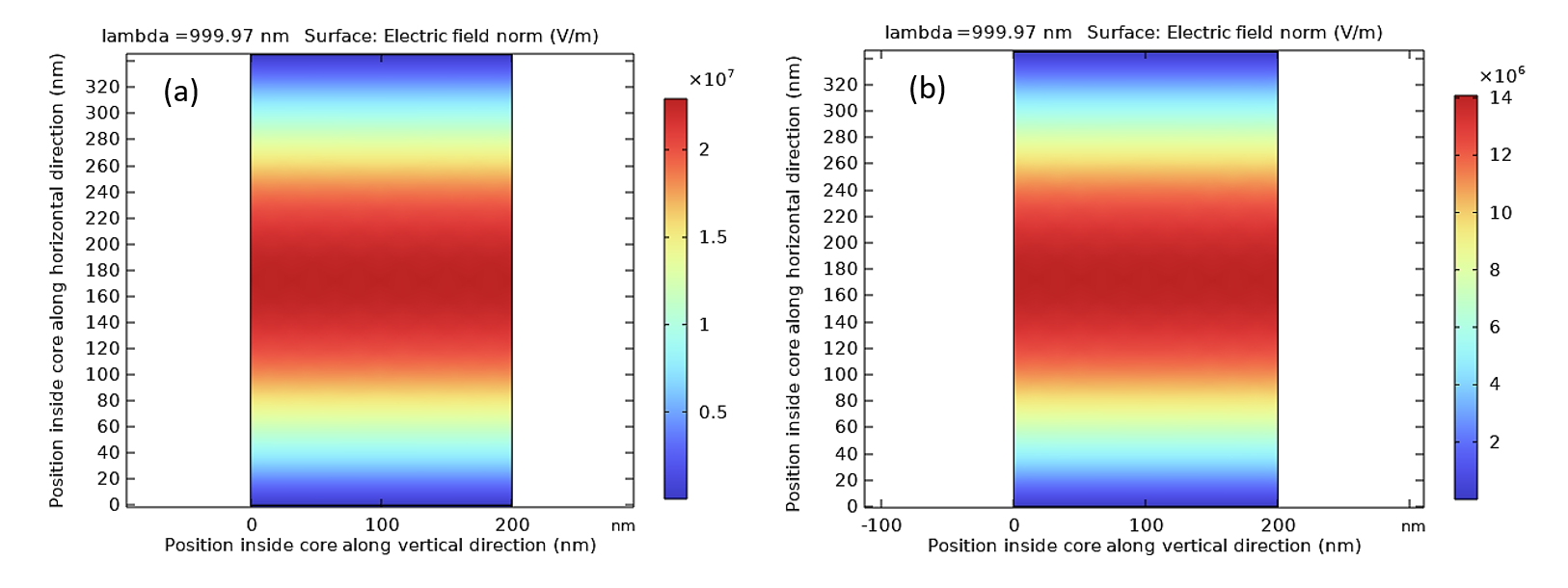}
  \caption{\label{fig:1} Mode profiles inside the cores of the (a) VM cavity and (b) BR cavity.}
\end{figure}

\begin{figure}
  \includegraphics[scale=0.7]
  {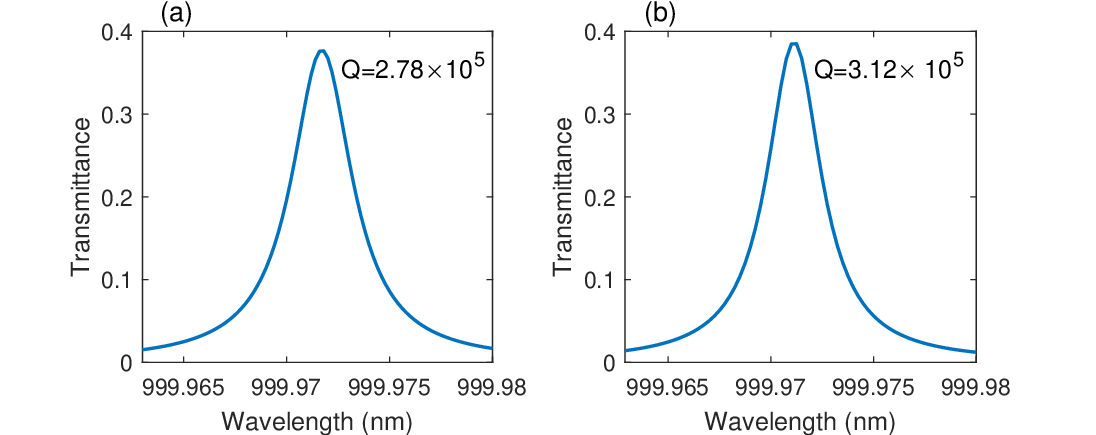}
  \caption{\label{fig:2} Spectral profiles of the transmission peaks through the (a) VM cavity and (b) BR cavity.}
\end{figure}

The transmission profiles of the VM and BR cavities are also shown in Figs. 2a and 2b, respectively. The similarity in cavity mode and transmission profiles justifies replacing BRs with VMs to mitigate convergence issues.

\noindent [S1] Malitson, Ian H. "Interspecimen comparison of the refractive index of fused silica." Journal of the optical society of America 55, no. 10 (1965): 1205-1209.

\noindent[S2] Gao, Lihong, Fabien Lemarchand, and Michel Lequime. "Exploitation of multiple incidences spectrometric measurements for thin film reverse engineering." Optics express 20, no. 14 (2012): 15734-15751.

\noindent[S3] Hagedorn, H., and J. Pistner. "High reflecting dielectric mirror coatings deposited with plasma assisted reactive magnetron sputtering." In Optical Systems Design 2015: Advances in Optical Thin Films V, vol. 9627, pp. 193-198. SPIE, 2015.

\end{document}